\documentclass{sig-alternate}

\usepackage[ruled,linesnumbered]{algorithm2e}
\usepackage{xcolor}
\usepackage{url}

\def\sharedaffiliation{\end{tabular}\newline\begin{tabular}{c}}

\usepackage{etoolbox}
\makeatletter
\patchcmd{\maketitle}{\@copyrightspace}{}{}{}
\makeatother

\begin{document}






%

\title{Building the Brazilian Academic Genealogy Tree 
}
%
%
%
%
%

\numberofauthors{4} 
%

\author{
	\alignauthor
       	\hspace{-1cm}
		 Wellington Dores\\
	\alignauthor
		\hspace{-4cm}
		Elias Soares\\
	\alignauthor
    	\hspace{-6cm}
		Fabr\'icio Benevenuto\\
	\alignauthor
     	\hspace{-6cm}
		Alberto H. F. Laender\\
	\sharedaffiliation
	\affaddr{Department of Computer Science}  \\
	\affaddr{Universidade Federal de Minas Gerais}   \\
	\affaddr{Belo Horizonte, MG, Brazil} \\
	\email{\{wellingtond, eliassoares, fabricio, laender\}@dcc.ufmg.br}
}


\maketitle
\begin{abstract}
Along the history, many researchers provided remarkable contributions to science, not only advancing knowledge but also in terms of mentoring new scientists. Currently, identifying and studying the formation of researchers over the years is a challenging task as current repositories of theses and dissertations are cataloged in a decentralized way through many local digital libraries. Following our previous work in which we created and analyzed a large collection of genealogy trees extracted from NDLTD, in this paper we focus our attention on building such trees for the Brazilian research community. For this, we use data from the Lattes Platform, an internationally renowned initiative from CNPq, the Brazilian National Council for Scientific and Technological Development, for managing information about individual researchers and research groups in Brazil.
\end{abstract}

%
%


%
%

%
%



\section{Introduction}

Science has evolved over the centuries as a system that not only promotes progress through the scientific method, but that is also centered on the processes of mentoring and teaching. The academic mentoring activity is a form of relationship that promotes the scientific development, as well as the formation and evolution of new researchers.
Despite the complex system behind science, most of the existing efforts in the literature that aim at measuring individuals' research productivity within a scientific community usually account only for the publications produced \cite{LimaSMSML15},  citations received~\cite{Benevenuto2016} and  collaborations established \cite{barabasi2002evolution,liu2005coauthorship}, neglecting the formation of new researchers.

There has been only a limited number of initiatives, by specific academic communities, in the sense of documenting, analyzing and classifying advisor-advisee relationships. Sometimes this kind of study considers a representation usually called academic genealogy tree~\cite{chang2003academic,david2012neurotree,jackson2007labor}, in which nodes represent researchers and relations indicate that a researcher was the advisor of another one. However, these efforts have focused on specific fields, such as Mathematics~\cite{jackson2007labor} and Neuroscience~\cite{david2012neurotree}, or have been restricted to a specific community as in the cases of a career retrospect of prominent American physicists~\cite{chang2003academic} and the tropical meteorology's academic community~\cite{hart2013family}. 
Although limited to specific locations and research areas, overall these efforts show that the analysis of such relationships in the form of a genealogy structure contributes to a greater understanding of a scientific community and of its individual values, allowing us to identify the impact generated by individuals in the formation of a community. For instance, Tuesta~\textit{et al.}~\cite{tuesta2014analysis} have analyzed the advisor-advisee relationship in the Brazilian exact and earth science field, correlating time and productivity throughout the advising relationship. Malmgreen \textit{et al.}~\cite{malmgren2010role} have investigated  mentorship fecundity using data from the Mathematics Genealogy Project.

Complementary to all these efforts, we have started an ambitious project towards building a large network that records the academic genealogy of researchers across fields and countries~\cite{dores2016genealogy}. Our preliminary work used data from NDLTD, the Networked Digital Library of Theses and Dissertations\footnote{http://www.ndltd.org}~\cite{FoxGMEAK2002ndltd}, and aimed to reconstruct advisor-advisee relationships from ETD records from many institutions around the world and from distinct disciplines. 

In this paper, we move one step forward by constructing academic genealogy trees from a completely different data source, the Lattes Platform\footnote{http://lattes.cnpq.br}.
Maintained by CNPq, the Brazilian National Council for Scientific and Technological Development, this platform is an internationally renowned initiative~\cite{lane2010nature} that provides a repository of researchers' curricula vitae and research groups, all integrated into a single system. All researchers in Brazil, from all levels (from junior to senior), are required to keep their curricula updated in this platform, which provides a great amount of information about the researchers' activities and their scientific production that can be used for many purposes. We then crawled the entire Lattes Platform and collected the curricula of all researchers holding a PhD degree. 
Next, we developed a basic framework to extract specific data from the collected curricula, identify and disambiguate the respective researchers, and establish their advisor-advisee relationships, from which we carried out a series of analyses that describe the main properties of the genealogy trees we were able to construct. Finally, we developed a first version of a system that allows users to browse and explore the academic genealogy trees. We believe that this is the first large-scale effort to generate a general academic genealogy tree involving as much distinct research fields as possible. We hope our framework can evolve into a much larger crowdsourcing system that stores a comprehensive collection of academic genealogy trees.

The rest of the paper is organized as follows. Next, we describe how we built our academic genealogy trees from the Lattes Platform. Then, we present a preliminary characterization of the academic genealogy trees we were able to built and discuss our findings. Finally, we conclude the paper and provide directions for future work.




\section{Building the Genealogy Trees}

In this section, we discuss how we built the researchers' individual academic genealogy trees (AGT's, for short) using data from the Lattes Platform. To build such AGT's, we first crawled the Lattes Platform and collected the curricula vitae (in XML format) of 222,674 researchers holding a PhD degree. Then, following the procedure described by Algorithm \ref{algorithm}, we parsed each collected curricula extracting the data required to build the researchers' AGT's. Such data appears basically in two specific sections of each curriculum: the \textsf{Identification} section, which includes the researcher's name, institution and degrees held, and the \textsf{Mentorships} section, which includes the researcher's list of all Master's and PhD students she has advised in her career. Note that the output of this procedure is actually a directed acyclic graph, since in her academic life a researcher might have had more than one advisor (e.g., PhD and Master's) or acted as a co-advisor for one or more students.   

Following Algorithm \ref{algorithm}, in order to build the AGT's, we first sort the set of all collected curricula according to the researcher's PhD degree year (line 1). This aims to establish a chronological order to build the individual AGT's, thus avoiding unnecessary name matchings when processing the advisees' curricula. Then we set the graph G empty (line 2).
Next, for each curriculum in the set C (lines 3 to 26), we execute the following three main steps: (i) search G for the respective researcher's node, creating a new node if it does not yet exist or updating it otherwise (lines 4 to 9); (ii) search G for the nodes of the researcher's PhD and Masters advisors, creating them if they do not yet exist or updating them otherwise, and then connect them to the researcher's node (lines 10 to 16); (iii) for each researcher's advisee, search G for her respective node, creating it if it does not yet exist or updating it otherwise, and then connect it to the researcher's node (lines 17 to 25).


A critical component of our algorithm is the search function present in lines 4, 10 and 17. Although the Lattes platform provides an internal identifier for each researcher with a registered curriculum, it is not always possible to use this mechanism to instantaneously identify another researcher whose name appears, for instance, in the list of mentorships of a specific curriculum. Thus, to overcome this problem, we have implemented a simple, but quite effective strategy to handle this typical name disambiguation problem~\cite{FerreiraGL12SigmodRec}, which considers the following parameters: the researchers' names, the names of their institutions, the titles of their theses or dissertations, and the respective years of defense. A detailed discussion of this name disambiguation strategy is out of the scope of this paper. However, it is worth noticing that, when connecting a researcher's node to the nodes of her advisors (lines 10 to 16), in most cases we use only her advisor's name and the name of the institution where she earned a degree to match the respective nodes.   



\begin{algorithm}[ht]
\SetKwInput{KwData}{Input}
\SetKwInput{KwResult}{Output} 

\KwData{A set C of Lattes Curricula;}
\KwResult{A graph G with all AGT's built;}

Sort C by the researchers' PhD degree year\;
Set G empty\;
 \ForEach{Curriculum c in C}
 	{Search G for the researcher's node n\;
  	\eIf{there is no such a node in G}
       	{Create node n\;}
   		{Update the academic attributes of n\;}
  	Search G for the nodes p and m of the researcher's PhD and Master's advisors\;
  	\eIf{either p or m are not found}
       	{Create them\;}
        {Update the academic attributes of p and m\;}
    Connect p and m to n\;
  	\ForEach{advisee in c}
       	{Search G for the advisee's node a\;
    	\eIf{there is no such a node in G}
           	{Create node a\;}
    		{Update the academic attributes of a\;}
    	Connect a to n\;}}
            
\caption{The AGT Bulding Procedure}
\label{algorithm}
\end{algorithm}


\vspace{-0.2cm}
\section{Characterizing the AGT's}

In this section, we briefly characterize some aspects of the AGT's we have been able to build. Our main motivation is to identify aspects that highlight the legacy of a researcher, measured in terms of formation of other researchers, and not in terms of the traditional counts of publications, impact factor, and scientific discoveries. 

\vspace{-0.2cm}
\begin{table}[ht]
\centering
\caption{Graph Characterization}
\label{tabGraph}
\begin{tabular}{|c|c|}
\hline
\# of Nodes              & 903,183     \\ \hline
\# of Edges              & 1,144,051   \\ \hline
\# of Trees              & 70,610      \\ \hline
\# of Components         & 22,061      \\ \hline
Avg. Tree Size           & 40.19       \\ \hline
Avg. Tree Width           & 3.81	       \\ \hline
\end{tabular}
\end{table}

Table~\ref{tabGraph} shows some figures about the AGT's. Besides basic figures such as number of nodes, edges and trees, the later defined by the number of ``roots" found in the graph (i.e., nodes without a known advisor), the  table also shows the number of components (i.e., connected trees) and the values of two important metrics: the average tree size and the average tree width. The values of these two last metrics are calculated by dividing, respectively, the number of descendants by the number of subtrees (average size) and the number of out-links of all nodes by the number of nodes (width).  

\begin{figure}[ht]
\caption{Cumulative Distribution Function of the Tree Sizes}
	\includegraphics[scale = 0.6]{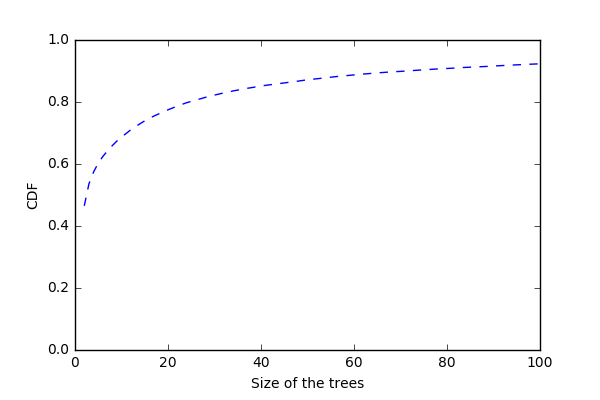}
    \label{sizes_cdf}
    \vspace{-0.2cm}
\end{figure}

\begin{figure}[!htb]
\vspace{-0.2cm}
\caption{Tree Depth Distribution}
	\includegraphics[scale = 0.6]{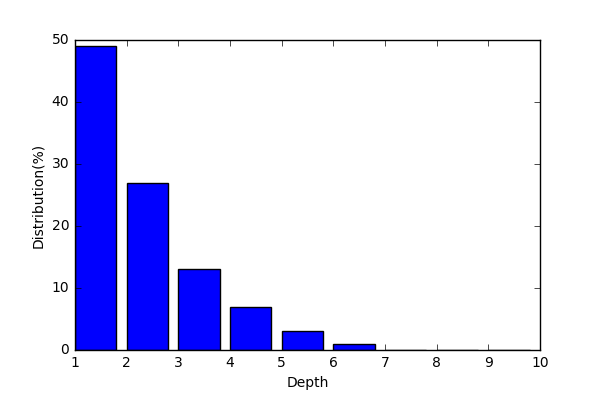}
    \label{fecundity_size}
\end{figure}

We have found in total 70,610 AGT's with 40.19 nodes on average. The average width of such trees is 3.81, i.e., each advisor in our dataset advised on average 3.81 PhD or Master's students. Despite the average size of the trees being 40.19, the 10 largest trees have more than 5,000 nodes, although 80\% of them have  less than 20 nodes, as shown by the graph in Figure~\ref{sizes_cdf}. On the other hand, almost half of the trees have depth 1, as shown in Figure~\ref{fecundity_size}. If we consider the width and the depth of a tree as its largest width and depth, respectively, we noted that trees are about 6.77 times wider than deeper in the Brazilian AGT's. This number is much higher  in comparison with the same ratio for trees built from NDLTD data~\cite{dores2016genealogy}, which is 2.48. We conjecture that this difference might be related to the quality of the trees we have obtained from both sources. NDLTD contains theses and dissertations from many institutions and countries, but it is unclear which scientific community it represents. On the other hand, Lattes represents an entire and complete scientific community, as basically all Brazilian researchers are forced to regularly update their academic records on the platform. We hope to incorporate many different data sources in our system and also allow users to fix and add their specific data, thus allowing one to better understand the idiosyncrasies from particular countries, research areas, or scientific groups, and their impact on scientific formation.

\vspace{-0.2cm}

We now comment on Table~\ref{tabCountries} that lists the six most important foreign countries where Brazilian researchers obtained a Master's or PhD degree. These six countries accounts for over 90\% of the Brazilian researchers who chose to study abroad. We note that Portugal appears in first place, which might be explained by the same language spoken in  both countries.  These results highlight how rich the data from Lattes is and the kind of findings we can exploit by deepening our analysis of the  AGT's built from them. 

\begin{table}[ht]
\centering
\vspace{-0.6cm}
\caption{The six most popular foreign countries from where Brazilian researchers earned a PhD or Master's degree}
\label{tabCountries}
\begin{tabular}{|c|c|c|}
\hline
\textbf{Country} & \textbf{PhD} & \textbf{Master's} \\ \hline
Portugal              & 1,179 & 300   \\ \hline
USA              & 891 & 254 \\ \hline
UK        &  853   & 219 \\ \hline
Spain              & 802    & 162 \\ \hline
France           & 660 &   248     \\ \hline
Argentina           & 584 &	 41      \\ \hline
\end{tabular}
\end{table}


\begin{figure*}[!htb]
\centering
\vspace{-0.1cm}
\caption{Example of an Academic Genealogy Tree Built from Lattes Data}
     \includegraphics[width=15cm, height=9.65cm]{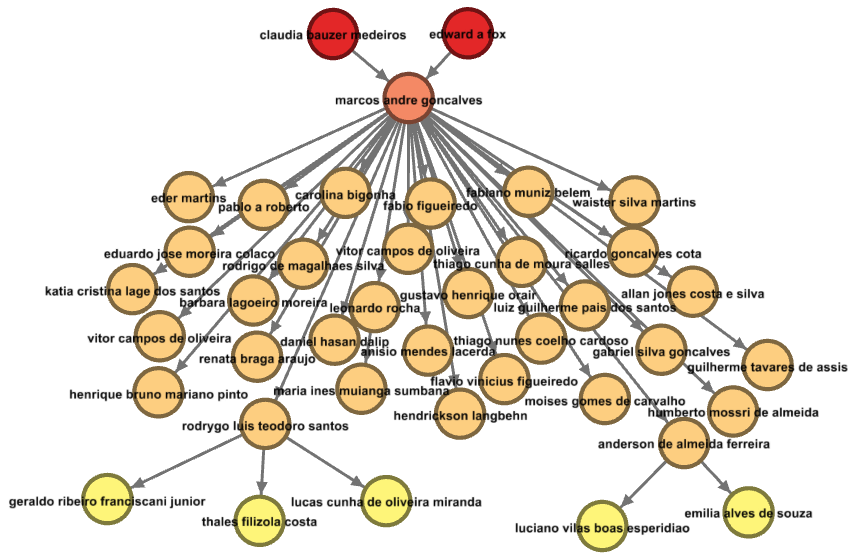}
    \label{redeExample}
\vspace{-0.3cm}   
\end{figure*}

\section{Conclusions and Future Work}

In this work, we used data crawled from the Lattes Platform to construct academic genealogy trees. Although still preliminary, our effort identified a number of interesting findings related to the structure of academic formation in Brazil, which highlight the importance of cataloging academic genealogy trees. Our effort, together with our previous work using data from the NDLTD~\cite{dores2016genealogy},  allowed us to identify many challenges that we need to tackle towards developing a large repository that records the academic genealogy of researchers across fields and countries. More importantly, we have developed a first version of a system 
that deploys the dataset studied here and allows users to browse the academic genealogy trees\footnote{\url{http://www.sciencetree.net}}. To briefly illustrate the potential of this system, Figure~\ref{redeExample} shows an excerpt of the genealogy tree of Dr. Marcos Andr\'e Gon\c{c}alves, a Brazilian associate professor from the Universidade Federal de Minas Gerais (UFMG), who is a well known member of the digital library community.  
 
The colors in the figure represent the levels in the AGT. The red nodes correspond to Dr. Gon\c{c}alves' advisors during his Master's (Prof. Claudia Bauzer Medeiros, from UNICAMP, Brazil) and PhD (Prof. Edward A. Fox, from Virginia Tech, USA) studies. The main subtree (the one rooted by an orange node) includes the graduate (Master's and PhD) students that have been advised by Dr. Gon\c{c}alves, which, in turn, span an additional level of subtrees (the yellow ones). Thus, by analyzing such a kind of tree we hope to be able to better understand a research lineage. More important, we believe this system represents a preliminary step towards the understanding of more important questions related to science, which we will be able to easily answer once we have a world-wide academic genealogy tree. For example, this system would allow us to identify the important researchers within areas and the role they have played on the creation and evolution of scientific communities, and even of novel fields. It would also provide a better understanding about where research areas came from, the birth and death of research communities, the identification of one's academic lineage, and the role of interdisciplinary formation on the evolution of specific research fields. Ultimately, it would allow us to better comprehend the evolution of science and consequently, of our society. We note, however, that our current version of the system is still beta and its development is part of our future work.

\section*{Acknowledgments}

This research is funded by grants from CAPES, CNPq and FAPEMIG. The last author was also supported by IEAT, the UFMG Institute for Advanced Transdiciplinary Studies.


%
\bibliographystyle{abbrv}

\end{document}